\documentstyle[aps,amssymb,prl,multicol]{revtex}

\begin{document}
\draft
\title{Antilocalization in a 2D Electron Gas in a Random Magnetic Field}
\author{D. Taras-Semchuk$^1$ and K. B. Efetov$^{1,2}$}
\address{$^1$ Theoretische Physik III, Ruhr-Universit\"{a}t\\
Bochum, Universit\"{a}tsstr. 150, 44780 Bochum, Germany\\
$^2$ L. D. Landau Institute for Theoretical Physics, 117940 Moscow, Russia}
\date{\today}
\maketitle

\begin{abstract}
We construct a supersymmetric field theory for the problem of a
two-dimensional electron gas in a random, static magnetic field. We find a
new term in the free energy, additional to those present in the conventional
unitary $\sigma $-model, whose presence relies on the long-range nature of
the disorder correlations. Under a perturbative renormalization group
analysis of the free energy, the new term contributes to the scaling
function at one-loop order and leads to antilocalization.
\end{abstract}


\vspace{-0.4cm}

\begin{multicols}{2}  

A question of long-standing controversy has been the influence of a random,
static magnetic field on a two-dimensional electron gas. This problem is
relevant to a wide range of physical situations, including the quantum Hall
effect in the half-filled Landau level \cite{KalHalp} and high-Tc
superconductivity in the gauge-field description \cite{IofNag}. Following
the application of a succession of analytical methods to this problem (e.g.
\cite{Alt}), Aronov {\em et al.} \cite{Aronov} constructed a $%
\sigma $-model for 2D electron sea in a delta-correlated random magnetic
field (RMF). They found that the $\sigma $-model was identical to the one
derived previously for disordered systems with broken time-reversal
invariance (unitary ensemble)\cite{Efetov},
and therefore concluded that all states are
localized as for the unitary ensemble. 
In particular, they did not find the additional
term in the $\sigma $-model proposed in Ref. \cite{arovas}; this term
appeared in Ref. \cite{arovas} as a result of an incorrect relation between
a correlator of Hall conductivities and the average longitudinal
conductivity. Although recently some arguments in
favor of delocalization by the RMF were presented \cite{miller}, it is
apparently commonly believed that all states of a 2D electron gas in the RMF
are localized and Ref.~\cite{Aronov} gives a proper description of the
model.

In this paper, we reconsider the problem presenting a more careful
derivation of a proper $\sigma $-model. It represents a generalization of
the supermatrix $\sigma $-model \cite{Efetov} to the case of a disorder
potential with long-range spatial correlations. For the RMF model, a new
term (different from the one of Ref.~\cite{arovas}) appears in the free
energy, additional to those present in the conventional unitary $\sigma $%
-model. The presence of this term is a consequence of the breaking of a
``gauge'' symmetry due to an anomaly at large momenta. The {\em long-range}
nature of the vector potential 
correlations corresponding to the RMF 
are crucial for our derivation. Under a perturbative renormalization
group (RG) analysis of the free energy, the new term contributes to the
scaling function at one-loop order and results in the
antilocalizing behaviour.

A related model which displays electron delocalization 
is the random flux model which describes, for example, electron
hopping on a bipartite lattice structure with link disorder (see e.g. \cite
{Gade,Altland}). Here a tendency towards
delocalization is displayed as the band center is approached
\cite{sugiya_etc,kalmey_etc}, due to the
existence of a chiral symmetry at the band center. 
However investigations of a lattice formulation of the RMF problem 
(equivalent to the random flux model at energies far from the band center)
remain controversial: conclusions are divided between the localization of
all states \cite{sugiya_etc} and the existence of a critical region
\cite{kalmey_etc}. The existence of a metal-insulator transition for
an RMF model with a spatially-correlated magnetic field has also been asserted
recently \cite{sheng}. It is
also of interest that, even for the case of one dimension, recent analytical 
\cite{Izrailev} and numerical \cite{Moura} work supports the existence of a
metal-insulator transition in the presence of sufficiently long-ranged
disorder correlations. It is well-known that spin-orbit impurities in 2D
also lead to antilocalization \cite{Hikami}, although by an entirely
different mechanism than that considered here.

In the following we consider the Hamiltonian, 
\begin{eqnarray}
{\cal H}=(-i\nabla 
-e{\bf A}/c)^{2}/(2m)-\epsilon _{F}.
\end{eqnarray}
A delta-correlated magnetic field leads to a vector potential, ${\bf A}$,
with a long-ranged and transversal correlator, 
$\langle A^{i}({\bf r})A^{j}({\bf r}^{\prime })\rangle  
=2 (mc/e)^2 V^{ij}({\bf r}-{\bf r}^{\prime })$ where
\begin{eqnarray}
\quad V_{{\bf q}}^{ij} &=&\nu _{F}^{2}\gamma \frac{1}{(q^{2}+\kappa
^{2}p_{F}^{2})}\left( \delta ^{ij}-\frac{q^{i}q^{j}}{q^{2}}\right) , 
\label{a2}
\end{eqnarray}
for some strength $\gamma $. Here $\epsilon _{F}$ and $v_{F}$
are the Fermi-energy and Fermi-velocity and $\kappa \ll 1$ is a cutoff that
renders finite the otherwise-infinite range of the disorder correlations. At
the end of calculations one should take the limit $\kappa \rightarrow 0$.
This cutoff appears \cite{Alt,Aronov} in the divergent single-particle
lifetime: for example, the simple Born approximation yields $\tau _{{\rm BA}%
}^{-1}=2\epsilon _{F}\gamma /\kappa $ while the self-consistent Born
approximation (SCBA) yields the less divergent $\tau _{{\rm SCBA}%
}^{-1}=2\epsilon _{F}(\gamma \ln (1/\kappa )/\pi )^{1/2}$. The inverse
transport time however remains convergent, $\tau _{{\rm tr}}^{-1}=\epsilon
_{F}\gamma $. We emphasize that our method remains valid for an arbitrary
choice of the correlator $V^{ij}({}{\bf r})$, with minor modifications for
scalar rather than vector potential disorder.

We now derive the free energy functional. Following standard methods and
notation \cite{Efetov}, an averaged product of Green's functions may be
expressed as a functional integral 
%
%
weighted by a Lagrangian that is quadratic in an eight-component,
supersymmetric $\psi $-field, 
\begin{equation}
{}{\cal L}=i\int \bar{\psi}({\bf r})\left( {\cal H}{}_{0}-\frac{ie}{mc}{}%
\tau_3{\bf A}.\nabla -\frac{e^{2}}{mc^{2}} A^{2}\right) \psi (%
{\bf r}{})d^{2}{\bf r}.  \label{Lag0}
\end{equation}
Here ${\cal H}{}_{0}=\epsilon +\epsilon _{F}+\nabla ^{2}/(2m)-\omega \Lambda
/2$, and $\tau _{3}$ and $\Lambda $ are the $z$-Pauli matrix in time-reversal
and retarded-advanced spaces. 
We average over ${\bf A}$ (neglecting the term in $A^{2}$ 
as it gives only a
slowly varying contribution to the chemical potential) 
to induce a term in the Lagrangian that is quartic in the $\psi $-fields
and, in contrast to the case of short-range impurities, non-local in
position. Decoupling this term via an 8$\times $8 matrix $Q({\bf r},{\bf r}%
^{\prime })$ and integrating over the $\psi $-fields, we come to the
Lagrangian 
\begin{eqnarray}
{\cal L}{} &=&\int \left[ -\frac{1}{2}{\rm Str}\ln \left( i{}{\cal H}_{0}+%
\hat{V}_{{\bf r,r}^{\prime }}\widetilde{Q}({}{\bf r},{}{\bf r}^{\prime
})\right) \right.   \nonumber \\
&&\left. \mbox{}\qquad +\frac{1}{4}{\rm Str}\left( Q({}{\bf r},{}{\bf r}%
^{\prime })\hat{V}_{{\bf r,r}^{\prime }}Q({}{\bf r}^{\prime },{}{\bf r}%
)\right) \right] d^{2}{\bf r}d^{2}{\bf r}^{\prime },  \label{Lag1}
\end{eqnarray}
with $\hat{V}_{{\bf r,r}^{\prime }}=-(1/2)\sum_{ij}V^{ij}({}{\bf r}-{}{\bf r}%
^{\prime })(\nabla_{{\bf r}}^{i}-\nabla 
_{{\bf r}^{\prime }}^{i})(\nabla _{%
{\bf r}}^{j}-\nabla_{{\bf r}^{\prime }}^{j})$. Also $\widetilde{Q}=Q_{\Vert
}+iQ_{\perp }$, where $Q=Q_{\Vert }+Q_{\perp }$ and $Q_{\Vert }$ ($Q_{\perp }
$) commutes (anticommutes) with $\tau _{3}$. 
The $Q$-matrix satisfies the standard symmetries $Q({}%
{\bf r},{}{\bf r}^{\prime })=\bar{Q}({}{\bf r}^{\prime},{}{\bf r})
=KQ^{+}({}%
{\bf r}^{\prime}
,{}{\bf r})K$, with $K$ defined as in Ref.~\cite{Efetov}.

To search for the saddle-point to Eq.~(\ref{Lag1}), we take $Q_{\perp }=0$
and $Q$ to depend on ${\bf r}-{\bf r}^{\prime }$ only. After Fourier
transforming, the saddle-point equation reads $Q_{{\bf p}}=g_{{\bf p}}$,
where 
\[
\left( \epsilon -\xi _{{\bf p}}-\frac{\omega }{2}\Lambda \right) g_{{\bf p}%
}-2i\int \frac{d^{2}{\bf p}_{1}}{(2\pi )^{2}}V_{{\bf p}-{\bf p}%
_{1}}^{ij}p_{1}^{i}p_{1}^{j}Q_{{\bf p}_{1}}g_{{\bf p}}=-i,
\]
and $\xi _{{\bf p}}={\bf p}^{2}/(2m)-\epsilon _{F}$. Writing $Q_{{\bf p}%
}=\Lambda _{{\bf p}}$, the required form for $\Lambda _{{\bf p}}$ is in
general complicated. The analysis simplifies however in the limiting cases
of $\kappa \gg (\epsilon _{F}\tau )^{-1}$ and $\kappa \ll (\epsilon _{F}\tau
)^{-1}$, where $\tau $ is the mean free time that is to be found from the
solution. The latter case is relevant here. In both cases we may take $%
\Lambda _{{\bf p}}=i/(\xi _{{\bf p}}+i\Lambda /(2\tau ))$, while keeping $%
\epsilon _{F}\tau \gg 1$. In the limit $\kappa \ll (\epsilon _{F}\tau )^{-1}$
one comes to the SCBA solution for the scattering lifetime, $\tau =\tau _{%
{\rm SCBA}}$.

To consider fluctuations around the saddle-point, we separate them into
three types: (a) hard massive, associated with the fluctuations of the
eigenvalues of the $Q$ and the cooperons, with the mass $\tau ^{-1}$ that is
very large in the limit $\kappa \rightarrow 0$, (b) soft massive, with the
mass $\tau _{{\rm tr}}^{-1}$ that remains finite, and (c) massless. The
condition $\epsilon _{F}\tau \gg 1$ ensures that fluctuations of type (a)
are irrelevant and so $Q^{2}({}{\bf r},{}{\bf r}^{\prime })=1$ and $[Q,\tau
_{3}]=0$. To obtain a functional for types (b) and (c), we write 
\begin{equation}
Q_{{\bf p}}({\bf R})=U\left( {\bf R}\right) Q_{{\bf p}}^{\left( 0\right) }%
\bar{U}\left( {\bf R}\right) ;\;Q_{{\bf p}}^{\left( 0\right) }=V_{{\bf n}}(%
{\bf R})\Lambda _{{\bf p}}\bar{V}_{{\bf n}}\left( {\bf R}\right) 
\label{param}
\end{equation}
with ${}{\bf R}=({}{\bf r}+{}{\bf r}^{\prime })/2$ and ${\bf n}={\bf p}/|p|$%
. In Eq. (\ref{param}), $U\left( {\bf R}\right) $ is independent of ${\bf n}$%
, and $Q_{{\bf p}}^{\left( 0\right) }\left( {\bf R}\right) $ contains only
non-zero harmonics in ${\bf n}$ around the Fermi surface. The supermatrix $%
Q_{{\bf p}}^{\left( 0\right) }\left( {\bf R}\right) $ can be parametrized, for
example, as 
\begin{equation}
Q_{{\bf p}}^{\left( 0\right) }\left( {\bf R}\right) =\Lambda _{{\bf p}%
}\left( 1+iP_{{\bf n}}\left( {\bf R}\right) \right) \left( 1-iP_{{\bf n}%
}\left( {\bf R}\right) \right) ^{-1}  \label{a6}
\end{equation}
with $\int d{\bf n}P_{{\bf n}}=0$, $\overline{P}_{{\bf n}}=-P_{{\bf n}}$, $%
\left\{ P_{{\bf n}},\Lambda \right\} =0$. Both $U\left( {\bf R}\right) $ and 
$P_{{\bf n}}\left( {\bf R}\right) $ vary slowly with ${\bf R}{}$, that is,
on length scales longer than $l=v_{F}\tau $.

The parametrization for $Q_{{\bf p}}\left( {\bf R}\right) $ guarantees that
the final free energy functional describing fluctuations will be invariant
under global rotations $U\left( {\bf R}\right) \rightarrow U_{0}U\left( {\bf %
R}\right) $ with $\bar{U}_{0}=U_{0}^{-1}$ independent on coordinates, this
invariance being present in the initial Lagrangian, Eq. (\ref{Lag0}).

As we are interested in the low-frequency behavior of transport coefficients
the massless modes are of greatest interest for us. However, we cannot simply
neglect the soft massive modes (b), and our aim now is to integrate 
them out, thus reducing the free energy to an effective functional containing
only $U\left( {\bf R}\right) $. Due to the soft mass of the (b) modes it is
sufficient to apply a Gaussian approximation in $P_{{\bf n}}\left( {\bf R}%
\right) $. Higher order terms give a small contribution provided the
inequality $\epsilon_F\tau _{{\rm tr}}\gg 1$ is fulfilled.

The crucial feature of this integration is that, 
as a consequence of the long-range correlations of the disorder, 
Eq.~(\ref{a2}),
it requires a cutoff at large momentum $k$ (short
distance) which {\em remains} in the final functional.  
There is some freedom in how to choose
the cutoff for fluctuations of $P_{{\bf n}}\left( {\bf R}\right) $
as the symmetry under the global
rotations is already guaranteed by the choice of parametrization
and there are no other symmetries to be
respected in the initial Lagrangians, Eq. (\ref{Lag0}, \ref{Lag1}). The
simplest way is to write this cutoff as the distance $l_{0}$, where $%
l_{\rm tr}=v_F\tau_{\rm tr}
\gg l_{0}\gg l$, below which the fluctuations are
suppressed 
\begin{equation}
P_{{\bf nk}}=0\text{ for }\left| {\bf k}\right| >l_{0}^{-1}  \label{a7}
\end{equation}
where $P_{{\bf nk}}$ are Fourier components of $P_{{\bf n}}\left( {\bf R}%
\right) =\sum_{{\bf k}}P_{{\bf nk}}\exp (i{\bf k}.{\bf R}\tau _{3})$. 
As we have to
calculate only Gaussian integrals over $P_{{\bf nk}}$, Eq. (\ref{a7}) is
very convenient for explicit computation. At the end of the calculation we
will take $l_{0}\rightarrow l$. 


We remark that to recover the conventional $\sigma $-model for short-range
(delta-correlated) impurities \cite{Efetov}, we may set $V_{{\bf n}}=1$ in
Eq.~(\ref{param}) so that $Q$ is a function of ${\bf R}$ only. It then
follows that the free energy possesses a local ``gauge'' 
invariance (LGI) under the transformation 
$U({}{\bf R})\rightarrow U({}{\bf R})h({}{\bf R})$, where $[h({}{\bf R}%
),\Lambda ]=0$ (note the term ``gauge'' is used here in a separate sense from
that of the original vector potential). 
It is the loss of this invariance in the case of
long-ranged disorder that leads finally to antilocalization. Before
explaining this point we present the expression for the free energy
functional $F[U]$ obtained after integration over the non-zero harmonics $P_{%
{\bf nk}}$:
\begin{eqnarray}
F[U] &=&\int d^{2}{\bf r}\left[ \frac{\pi \nu }{8}{\rm Str}\left[ D(
\nabla 
Q\left( {\bf r}\right) )^{2}+2i\omega \Lambda Q\left( {\bf r}\right) \right]
\right.   \nonumber \\
&&\left. \hspace{1.5cm}-\frac{\beta }{32\pi }\left[ {\rm Str}(\tau _{3}\Lambda 
\bar{U}\left( {\bf r}\right) \nabla U\left( {\bf r}\right) )\right] ^{2}%
\right] ,  \label{action}
\end{eqnarray}
where $D=v_{F}^{2}\tau _{{\rm tr}}/2$ is the classical diffusion
coefficient, $Q\left( {\bf r}\right) =U\left( {\bf r}\right) \Lambda \bar{U}%
\left( {\bf r}\right) ${\bf , }and $\beta =\tau _{{\rm tr}}/\left( 2\tau
\right) \gg 1$.

The first line in Eq. (\ref{action}) corresponds to the conventional $\sigma 
$-model for the unitary ensemble that predicts localization. This part has been
derived in Ref. \cite{Aronov} for the model in question. The second line in
Eq. (\ref{action}) is the main result of the present paper. The first line
respects the LGI but the term in the second line violates this
invariance and leads eventually to antilocalization.

At first glance, the violation of the LGI seems impossible
because, on 
replacing $U({}{\bf R})\rightarrow U({}{\bf R})h({}{\bf R})$ in Eqs.
(\ref{Lag1}, \ref{param}), we can change the variables of integration $P_{%
{\bf n}}\left( {\bf R}\right) \rightarrow \overline{h}\left( {\bf R}\right)
P_{{\bf n}}\left( {\bf R}\right) h\left( {\bf R}\right) $ when integrating
over the non-zero harmonics. This would remove $h\left( {\bf R}\right) $
from the final expression for $F[U]$. However, this change of the variables
would replace Eq. (\ref{a7}) by a more complicated expression for the
cutoff. The LGI is therefore guaranteed only if the cutoff is
not important, that is, when
no ultraviolet divergencies appear in the integral over the non-zero
harmonics. 

Now let us explain the most important steps of the derivation of Eq. (\ref
{action}). First, we perform an expansion of the logarithm in the Lagrangian
(\ref{Lag1}), in both $\bar{T}_{{\bf n}}\nabla 
T_{{\bf n}}$ (where $T_{{\bf n}%
}=UV_{{\bf n}}$) and the collision integral (deviation of 
$\hat{V}_{{\bf r,r}^{\prime }}\widetilde{Q}({}{\bf r},{}{\bf r}^{\prime
})$
from its saddle-point value). 
If we keep
to first order in both of these terms, we recover the Muzykantskii and
Khmelnitskii $\sigma $-model with ballistic disorder \cite{Muz}. Although
such an expansion is sufficient to describe a long-ranged potential (for
which $\tau _{{\rm tr}}\gg \tau $), for the general case including a
short-range potential we require the corrections to the free energy in Ref.~ 
\cite{Muz} obtained by continuing this expansion to second-order in both
terms. The resulting terms in the free energy then complete the list of
those that contribute to Gaussian order in $P_{{\bf n}}$.

We find it convenient to Fourier transform from $P_{{\bf nk}}$ to angular
harmonics, $P_{{\bf nk}}=\sum_{m}P_{m{\bf k}}\exp \left( im\phi \tau
_{3}\right) $. Expressing the collision integral in terms of these
harmonics, we find that, as well as the SCBA scattering lifetime described
previously, a whole series of lifetimes associated with successive harmonics
appears in the free energy. We define the $m$th lifetime $\tau^{(m)}$ by 
\[
\frac{\Lambda }{2\tau ^{(m)}}=2\int \frac{d^{2}{\bf p}_{1}}{(2\pi )^{2}}V_{%
{\bf p}-{\bf p}_{1}}^{ij}p_{1}^{i}p_{1}^{j}\Lambda _{p_{1}}\cos (m\varphi ),
\]
where $\varphi $ is the angle between ${\bf p}$ and ${\bf p}_{1}$, so that $%
\tau ^{(0)}$ coincides with $\tau $ and $1/\tau _{{\rm tr}}=1/\tau -
1/\tau^{(1)}$. 
The free energy becomes $F=F_{0}+F_{\shortparallel }+F_{\perp
}+F_{{\rm unit}}$ with 
\begin{eqnarray*}
\hspace{-0.5cm}F_{0} &=&\pi\nu \hspace{-0.05cm}\int 
\frac{d^{2}{\bf k}}{(2\pi )^{2}}%
\sum_{m=1}^{\infty }{\rm Str}\left[ P_{m{\bf k}}P_{m,-{\bf k}}\frac{
\tau ^{(m)}%
}{\tau }\left( \frac{1}{\tau }-\frac{1}{\tau ^{(m)}}\right) \right.  \\
&&\left. \hspace{-0.9cm}+\frac{iv_{F}}{2}\Lambda \left( P_{m{\bf k}}P_{-m-1,-%
{\bf k}}\bar{k}^{\ast }+P_{m+1,{\bf k}}P_{-m,-{\bf k}}\bar{k}\right) \right]
, \\
F_{\shortparallel } &=&\pi \nu v_{F}\int d^{2}{\bf r}\sum_{m=1}^{\infty }%
{\rm Str}\left[ \Phi _{x}^{\shortparallel }\tau _{3}\Lambda \left(
P_{m}P_{-m-1}\right. \right.  \\
&&\left. \left. +P_{m+1}P_{-m}\right) -i\Phi _{y}^{\shortparallel }\Lambda
(P_{m}P_{-m-1}-P_{m+1}P_{-m})\right] , \\
F_{\perp } &=&\frac{-i\pi \nu v_{F}}{2}\int d^{2}{\bf r}\,{\rm Str}\left[
\Phi _{x}^{\perp }\tau _{3}\Lambda (P_{1}+P_{-1})\right.  \\
&&\left. \hspace{4.2cm}+i\Phi _{y}^{\perp }\Lambda (P_{1}-P_{-1})\right] ,
\end{eqnarray*}
\vspace{-0.8cm} 
\begin{equation}
\hspace{-1.8cm}F_{{\rm unit}}=\frac{\pi \nu }{8}\int d^{2}{\bf r\,}{\rm Str}%
\left[ D_{0}(\nabla Q)^{2}+2i\omega \Lambda Q\right] ,  \label{a10}
\end{equation}
where $\nu $ is the density of states, $\bar{k}=k_{x}+ik_{y}\tau _{3}$, $%
\Phi\left( {\bf r}\right) =\bar{U}\left( {\bf r}\right) \nabla
U\left( {\bf r}\right) $, and $\Phi^{\shortparallel }$ 
$(\Phi^{\perp })$
are the components of $\Phi$ 
that commute (anti-commute) with $\Lambda $.
Also $Q=U\left( {\bf r}\right) \Lambda \bar{U}\left( {\bf r}\right) $ and $%
D_{0}=v_{F}^{2}\tau /2$. As an intermediate step we have rescaled $%
P_{m}\rightarrow P_{m}\tau ^{(m)}/\tau $.

Having derived the free-energy functional for the fluctuations in $Q_{{\bf n}%
}\left( {\bf r}\right) $, Eq. (\ref{a10}), the next step is to integrate
over the non-zero harmonics, represented by $P_{{\bf n}}$, to derive a
functional in terms of only $U\left( {\bf r}\right) $. We may check
immediately that for short-range (delta-correlated) disorder, the
conventional unitary sigma-model is recovered: in this case, $\tau
^{(m)}\rightarrow \infty $ for $m\neq 0$, the $P_{m}$ fields become infinitely
massive and only the terms in $F_{{\rm unit}}$ remain as required. For the
case of the delta-correlated RMF, we find instead by
explicit evaluation that $1/\tau -1/\tau ^{(m)}=(2m-1)/\tau _{{\rm tr}%
}$, and $\tau ^{(m)}/\tau \simeq 1$, for $m\ll \kappa ^{-1}$. We then
average the terms in $\Phi$ 
with respect to the functional $F_{0}$. The
leading contribution, containing no more than two gradient operators, comes
from the second-order cumulant of $(F_{\shortparallel }+F_{\perp })$, 
according to
$F_{0}+F_{\shortparallel }+F_{\perp }\rightarrow -(1/2)\langle
(F_{\shortparallel }+F_{\perp })^{2}\rangle _{F_{0}}$.

To perform the averaging we apply what amounts to a straightforward
generalization of the standard contraction rules \cite{Efetov} to include
the angular harmonic space (see e.g. \cite{Aleiner}). The contribution of
the mixed term $F_{\perp }F_{\shortparallel }$ vanishes. The terms in $%
F_{\perp }$ lead to a dressing of the bare diffusion coefficient, $D_{0}$,
appearing in $F_{{\rm unit}}$: $D_{0}\rightarrow D=v_{F}^{2}\tau _{{\rm tr}%
}/2$. Taking into account only the contributions from $F_{{\rm unit}}$ and $%
F_{\perp }$ 
corresponds to the calculation of Ref. \cite{Aronov} and 
gives
the first line of Eq. (\ref{action}). The terms in $F_{\shortparallel }$
induce the new term in the free energy 
$F[U]$ (second line in Eq. (\ref{action}%
)), where 
the dimensionless coefficient $\beta $ can be written as
\begin{equation}
\beta =2\pi \int_{\left| {\bf s}\right| <s_{0}}\frac{d^{2}s}{\left( 2\pi
\right) ^{2}}\left( \frac{\partial ^{2}}{\partial s_{x}^{2}}+\frac{\partial
^{2}}{\partial s_{y}^{2}}\right) {\rm Tr}\ln \hat{L}\left( \bar{s}\right) 
\label{a11}
\end{equation}
Here ${\bf s=k}l_{{\rm tr}}/2$ is a rescaled momentum variable 
which must be cut off at some upper limit set by $l_{0}$ as
discussed above: we have $s_{0}=l_{{\rm tr}}/(2l_{0})\gg 1$. The
semi-infinite, tridiagonal matrix $\hat{L}$ in Eq. (\ref{a11}) has the
entries 
\[
\left( \hat{L}(\bar{s})\right) _{m,n}=(2m-1)\delta _{m,n}+i\bar{s}\,\delta
_{m+1,n}+i\bar{s}^{\ast }\,\delta _{m,n+1}
\]
for $1\leq m,n\ll \kappa ^{-1}$ and $\bar{s}=s_{x}+is_{y}$. 
We see that the integral in Eq. (\ref{a11}) is determined only by a
contribution of the boundary $\left| {\bf s}\right| =s_{0}$
(${\rm \det }\,\hat{L}$ depends on $s=|%
{\bf s}|$ only):
\begin{equation}
\beta =\left[ s\frac{d}{ds}\left( {\rm \ln }\,{\rm \det }\hat{L}(\bar{s}%
)\right) \right] _{s=s_{0}}.  \label{a13}
\end{equation} 
In the case where the cutoff $l_0$ is unimportant, 
such as for sufficiently short-range disorder, this term becomes
identically zero and the LGI is preserved as required. In the more
general case, we see that $\beta$ 
is determined by 
the asymptotes of the function $\det \hat{L}%
\left( \bar{s}\right) $, that in turn are 
determined by the range of the
disorder correlations. 
 
For the RMF model, we have 
succeeded in calculating the determinant of $\hat{L}$ (for $\kappa
\rightarrow 0)$ analytically as follows. If we set $\hat{L}_{N}$
as the truncation of $\hat{L}$ to $1\leq m,n\leq N$ and introduce $M_{N}(s)=%
{\rm \det }\hat{L}_{N}(s)/((2N-1)!!)$, we find the recurrence
relation 
\[
M_{N}=M_{N-1}+\frac{s^{2}}{(2N-1)(2N-3)}M_{N-2},
\]
with the boundary conditions $M_{0}=M_{1}=1$. Next
we introduce the function $%
f(z,s)=\sum_{N=0}^{\infty }M_{N}(s)z^{N}$ and derive from the recurrence
relation
a second-order differential equation for $f(z,s)$. As a result we come to the
simple formula 
\begin{eqnarray}
\lim_{N\rightarrow \infty }M_{N}(s)=\cosh (s)
\end{eqnarray}
and hence $\beta (s_{0})=s_{0}\tanh (s_{0})\simeq s_{0}\gg 1$, which should
be used in Eq. (\ref{action}).

We now perform a perturbative RG analysis on the free energy (\ref{action})
(see Ref.~\cite{Efetov}): we separate $U$ into slow ($\tilde{U}$) and fast ($%
U_{0}$) components, $U=\tilde{U}U_{0}$, and expand $U_{0}$ perturbatively in 
$P$, where $U_{0}=(1-iP)^{1/2}(1+iP)^{-1/2}$. 
As for the conventional
unitary $\sigma $-model, the diffusion coefficient $D$ receives a negative
correction at two-loop order. 
To derive the correction it is necessary to go below two
dimensions and analytically continue the finite result to 2D. 
For the free
energy (\ref{action}), the diffusion coefficient gains a further, positive
correction originating from the new term in $\beta $, this time at only
one-loop ($O(P^{2})$) order. At the same time, $\beta $ is not renormalized
in this order. The scaling function for $t=8/(\pi \nu D)$ in 2D becomes 
\begin{equation}
\frac{dt}{d\ln \lambda }=\frac{1}{2}t^{3}(\beta -1)+O(t^{4}),  \label{a15}
\end{equation}
where $\lambda $ is the running momentum cutoff for the $P$-fields, with the
solution 
\begin{equation}
t(\omega )=t_{0}\left[ (1+(\beta -1)t_{0}^{2}\ln (1/\omega \tau _{tr})\right]
^{-1/2}.  \label{a16}
\end{equation}

Finally we take the cutoff $l_{0}\rightarrow l$. If we also cut off the
range of the disorder correlations at the system size, $L$, we have $\kappa
\sim 1/(p_{F}L)$. This gives $\beta =\tau _{{\rm tr}}/2\tau \sim
(\ln (p_{F}L)/\pi
\gamma )^{1/2}\gg 1$. We see that the resistivity decreases as the frequency 
$\omega $ is lowered and eventually vanishes, corresponding to
antilocalization. 

In deriving the above scaling function it is crucial that the 
function $%
\beta (s_{0})$ for $s_{0}\rightarrow \infty $ survives dimensional
regularization below 2D.
In fact, the
exponential dependence of ${\rm \det }\,\hat{L}(s)$ on $s$ for large $s$
ensures that $\beta $ remains finite for any dimension $%
d>1$. For any finite range disorder correlations the function $\det \hat{L}%
\left( s\right) $ would grow algebraically. Extending Eq. (\ref{a11}, \ref
{a13}) to $d<2$, we see that $\beta $ vanishes in this case and the
conventional unitary $\sigma $-model is recovered.


In summary, we have considered the problem of a 2D electron gas in a
random, static magnetic field and found that the resistivity vanishes in the
limit of small frequencies, a behaviour which 
may be classified as antilocalization.
This result may have far-reaching consequences for gauge field models in 
the theory
of strongly correlated systems and lead to other interesting applications. 

We are grateful to I. Aleiner, A.Altland, V.E. Kravtsov, and A.I. Larkin for
many helpful discussions and gratefully acknowledge the financial support of
the {\it Sonderforschungsbereich} 237.

\end{multicols}
\end{document}